\documentstyle[12pt]{article}
\newcommand{\bibi}{\bibitem}

\newcommand{\vsp}{\vskip 5mm \noindent}

\newcommand{\half}{\frac {1}{2}}

\newcommand{\beq}{\begin{equation}}
\newcommand{\eeq}{\end{equation}\noindent}

\newcommand{\beqr}{\begin{eqnarray}}
\newcommand{\eeqr}{\end{eqnarray}\noindent}

\newcommand\up{\uparrow}

\newcommand\down{\downarrow}
\begin{document}

\title{ CHARGE-SPIN SEPARATION IN 2D FERMI SYSTEMS:
SINGULAR INTERACTIONS AS MODIFIED COMMUTATORS, AND SOLUTION OF
2D HUBBARD MODEL IN BOSONIZED APPROXIMATION } %%%%%%%%

\author{ P.W. Anderson and D. Khveshchenko \\
Joseph Henry Laboratories of Physics \\
Jadwin Hall, Princeton University \\
Princeton, NJ 08544* \\
}

\date{}
\maketitle
\begin{abstract}

The general 2-dimensional fermion system with
repulsive interactions (typified by the Hubbard Model) is
bosonized, taking into account the finite on-shell forward
scattering phase shift derived in earlier papers.  By taking this
phase shift into account in the bosonic commutation relations a
consistent picture emerges  showing the charge-spin separation and
anomalous exponents of the Luttinger liquid.

\end{abstract}

\vsp

$ ^*$ This work was supported by the NSF, Grant \# DMR-9104873

\pagebreak

The proper description of the effect of finite forward-scattering
on-shell phase shifts on Fermi systems for $D>1$ \cite{one}
has been the subject of a number of papers \lbrack 2-4\rbrack.
The existence of
such scattering, leading to on-shell singularities in the
$T$-matrix, was confirmed by Fukuyama and Narikiyo \cite{five},
di Castro, Castellani and Metzner \cite{six} and others. The
discovery which these papers confirmed followed from the fact
that when two particles are embedded in their respective Fermi
seas, effectively all soft recoils are forbidden to them by the
exclusion principle.  Under these conditions, the logarithm of
the $S$-matrix for relative motion retains a finite eigenvalue
$\eta_0$ in 2 dimensions,
$$S_0(Q)={\rm exp}\,i[\eta_0(Q)]$$
in the limit that the relative momentum $Q=k_\up-k'_\down\to 0$,
and that both states are below the Fermi level. (On general
grounds, it seems likely that this can happen in 3D as well, but
this has not been proven.) It is important to be clear that $S$ is
not the conventional ``T-matrix'' defined for a hole at $k$
scattering against another at $k'$, whose imaginary part
represents an incoherent decay process, vanishing at the Fermi
surface; $S$ is the on-shell scattering matrix of particles
embedded in the Fermi sea and its phase determines the boundary
condition for their asymptotic wave-functions at the origin of
relative coordinates $r-r'=0$.  Addition of a particle modifies
the wave functions of all other particles, and our endeavor is
to investigate the consequences of this fact.

Our initial description \cite{seven} was heuristic, merely pointing
out that the effect of such a phase shift mimics that of a change
in statistics by enforcing a partial exclusion principle between
electrons of opposite spins.  We also described the type of
singular interaction which would give energy shifts similar to
those which take place, and emphasized that these would ``trash''
Fermi liquid theory.  Some of the papers have focused on this
singular interaction (specifically Randeria, et al\cite{two} and
Stamp et al.\cite{three}.  The treatment in terms of an
interaction is in several respects unsatisfactory, as clarified
by Baskaran\cite{four}.  But even Baskaran's discussion does not
give us a clear insight into clean formal ways to deal with the
situation.

The problem is that the effect is best thought of as a
$\rm\underline {constraint}$ on the wave-functions, not as an
interaction.  This is most clearly seen in the Hubbard model,
where the effect of a strong enough repulsive potential $U\to\infty $
is to enforce a projective constraint, expressed as the
Gutzwiller projector acting on the kinetic energy in the $t-J$
model, for instance.  Since the exchange term also is expressible
purely in terms of projected operators, the $t-J$ system is
confined to the subspace defined by projected operators.

It is worth emphasizing that renormalization group derivations of
Fermi liquid theory as a theory of the low-energy states, such as
that of Shankar, implicitly assume a free Fermion starting
Hamiltonian.  If the starting problem itself is projected onto a
subspace, this property will remain after renormalization and
F.L.T. changes into the theory we shall derive.

In general (in 2D) the constraint appears as a phase shift, which
is a boundary condition for the asymptotic wave function in the
relative coordinates of a pair of particles.  Such a wave
function is indeterminate unless it has a boundary condition both
at $r-r'\to \infty$ and at $r-r'\to 0$  Arguments in several of
the original papers show that the rest of the particles may be
satisfactorily dealt with by taking the exclusion principle into
account, and multiparticle  encounters are not crucial.

This local boundary condition on the asymptotic wave function at
$r-r'\to 0$ is a kinematic, rather than a dynamic, effect: there
is a change in the wave functions of the particles, not directly
in their energy.  We are used to this with hard core potentials:
the effect is best expressed as one purely on the kinetic energy, not
on the potential.  This kinematic effect dominates here because
the scattering region where the potential acts is small, of order
$N^{-1}$ compared to the asymptotic region in which the kinetic energy
is modified.  The way to make this point is that such a
boundary condition can actually change the dimensionality of the
Hilbert space of allowed wave functions.  In simpler terms, such
a boundary condition forces the wave function's nodes to shift in
such a way that $\rm\underline { a}$ particle moves into or out
of the distant boundary, so that the same volume contains
$N\pm \eta/\pi$ particle states rather than $N$.  This is what is meant
by a change in the dimensionality of Hilbert space.  This change of Hilbert
space occurs in 1D even as a consequence of an ordinary
interaction potential (hence the flexibility of statistics in 1D)
but in all other dimensions it is distinct from the kind of
interaction effects which can be treated perturbatively.

The conclusion we came to is therefore that the effect of a finite
phase shift is best modelled as a modification of the algebra of
the particles, expressed in their commutation relations.
Projected Fermions
\beq (c^+_{i\sigma})_{proj}=P^i_G\,c^+_{i\sigma}=(1-n_{i-\sigma})
c_{i\sigma}^{\dag} \eeq
do not have the same commutation relations as ordinary Fermions,
obviously, but we have not found the Fermion representation
convenient to work with.  It is much simpler to use the bosonized
representation in terms of the Fermi surface\footnote{*The
bosonized version of Fermi liquid theory can be equivalently
thought of as the appropriate gauge theory in the presence of a
Fermi surface, since the bosonic variable is essentially the phase of
the Fermi surface wave function.} fluctuations \cite{eight,nine}.

Haldane, particularly, has emphasized that the most useful
description of the dynamics of a Fermi system is via the
operators $\Delta k_F$ describing the position of the Fermi
surface in $k$-space, taken to be dynamical variables, functions
of a coarse-grained space and time.  That is, he argues that
Luttinger's theorem holds exactly during sufficiently
long-wavelength and low-frequency fluctuations. (Parenthetically,
even the conventional derivations of Luttinger's
theorem\cite{ten} depend not on the convergence of
perturbation theory but merely on the assumption that excitations
precisely at the Fermi surface do not decay, hence the Green's
function is real.) We define operators
$$\Delta k_{F\sigma}(\Omega, r, t)$$
giving the Fermi surface fluctuations of spin $\sigma$ at a point
on the F.S. parametrized by $\Omega$, and at coarse-grained
$r,t$.  These are the bosonic variables: they commute for
different $\Omega$ and $r$, and, for non-interacting electrons,
for different $\sigma$.  We can introduce a phase variable
$\theta_\sigma$ of the wave function at the Fermi surface, which
is a function of $\Omega, r$ and $t$, and then $\Delta
k_{F\sigma}$ is
\beq \Delta k_{F\sigma}=\hat {n}_\Omega\cdot
\nabla\theta_\sigma \eeq
where $\hat {n}_\Omega$ is the local normal to the fiduciary
Fermi surface.  $\theta$ and $\Delta k_F$, which is equivalent to
the particle density at $\Omega, \rho\,(\Omega)$, are conjugate
variables, and have for free fermions canonical commutation
relations:
\beq [\theta, \rho]=i\pi\, \delta\, (r-r')\ \delta \,(\hat
\Omega-\hat{\Omega'}) \eeq
As Haldane has pointed out, this representation can be motivated
by the idea of expressing the Fermion field in terms of two real
operators $\rho$ and $\theta$
\beq \psi(x)=\rho\, e^{i\theta(x)}\eeq
rather than by the earlier ``Tomonaga'' definition of $\rho(q)$ as
a density of Fermions $\sum_k c^+_{k+q}\, c_k$.  This latter
representation is not possible when the Fermions are projected
operators. But we can still speak of a Fermi surface and a Fermi
surface phase for each spin which satisfies Luttinger's theorem,
hence determines the density of particles at each point on the
Fermi surface.
In this transcription of the original idea of bosonization we
follow Khveshchenko \cite{tenprime}.  If a Fermi surface exists this implies
zero-frequency modes at each point on it, hence separate,
independent conservation of particle and spin currents at the
Fermi surface at each $\Omega$ even allowing for Fermi surface
fluctuations, which may be integrably singular at low
frequencies.

However, this does not imply that, in the presence of
interactions, $\theta_\sigma$ and $\rho_\sigma$ (or $\Delta
k_{F\sigma}$) remain the appropriate canonically conjugate
variables.  These are variables which measure, respectively, the
particle number at a particular patch on the Fermi surface and
given spin and the phase of the wave function at the Fermi
surface.  If there is a finite phase shift for forward scattering
of opposite spin electrons, as we have
shown\cite{one,seven}, the order
of doing these operations matters.  If we add a particle of up
spin, the phase of the down spin wave function depends on
whether the particle of up spin was added before or after the
phase was measured.  The failure of commutation for opposite
spins is the phase shift $\eta/\pi$, just as adding a particle of
up spin below the Fermi surface enforces a change in up-spin
phase by the amount $\pi$.  We may express this by writing the
free particle commutator in matrix form:
\beq [\rho_\sigma,\ \theta_{\sigma'}]_{\rm bare}=i\pi
\left(\matrix{1& 0\cr
              0& 1\cr}\right)\,\delta (r-r')\delta(\Omega-\Omega')
\eeq
while
\beq
[\rho_\sigma,\ \theta_\sigma]_{\rm interacting}=i\pi
\left(\matrix{1&{\eta\over \pi}\cr
                 {\eta\over\pi}& 1\cr}\right)
\times \delta (r-r')\
\delta(\Omega-\Omega').\eeq

Let us explain these equations in detail. (5) means in the
one-dimensional model that if we insert an extra particle
into the Fermi sea at a point $r$, because of the exclusion
principle the wave-function at the Fermi surface (which is the
basic interpretation of equation (4)) must have an extra node
inserted into it near $r$, hence the phase difference between left and
right-going (or in---and out---going) waves must shift by $\pi$
as a consequence.  Hence after we insert one particle in
$\rho,\theta$ will change by $\pi$, but not vice versa: one is the
generator of displacements of the other.
(6) must be interpreted in exactly the same way.  The insertion of an
up-spin particle at $r$, near $\Omega$, means that the phase of
the down spin wave at $\Omega$ is shifted by $\eta$, while the
up-spin wave is shifted by $\pi$.  This means that $\theta_\up,
\rho_\up$ and $\theta_\down, \rho_\down$ are no longer canonically
conjugate; the correct canonically conjugate variables are
proportional to
\beq \theta _s = \frac{\theta _{\up} - \theta _{\down}}{\sqrt {2}}
\quad \rho_s = \frac{\rho_\up-\rho_\down}{\sqrt{2}}
\eeq and
\beq \theta_c = \frac{\theta_\up+\theta_\down}{\sqrt{2}}
\quad \rho_c = \frac{\rho_\up+\rho_\down}{\sqrt{2}}.\eeq

The equations of motion of the charge and spin bosons follow from
the commutation relations and the Hamiltonian, which as we
explained is simply the original kinetic energy, the interaction
terms being completely subsumed in the C.R. The Hamiltonian is
as for free particles the one given by Haldane,
$$ {\cal H} = \half \int d\,\Omega\int
d^D\, r\,v_F\,(\Omega)[\Delta k_{F}(\Omega), r,t)]^2$$
\beq {} = \half
\int d\,\Omega\sum_q
v_F[\rho^2(q,\Omega)+q^2\theta^2(q,\Omega)]. \eeq
Then
\beq [H,\theta_{c,s}(q,\Omega)]=v_{c,s} q\theta_{c,s}\, (q,\Omega)\eeq
with
$$  v_s = v_F(\Omega)(1-{\eta\,(\Omega)\over  \pi})$$
\beq v_c  = v_F (\Omega)\Big (1+ {\eta \,(\Omega)\over \pi}\Big )\eeq
and bosons are left as harmonic oscillator variables
with frequencies
$\left\{\matrix{qv_c&(\Omega)\cr
                qv_s&(\Omega)\cr}\right\}$

%%%%$\{qv_c\,(\Omega)\atop{qv_s(\Omega)}\}$

For free particles the Fermion operator is made up from bosons
via the formula
$${\rm Free:}\ \ \psi^\Omega_\sigma(r)=\rho_0e^{-{i\over
\sqrt{2}}(\theta_c+\sigma\theta_s)}\eqno(12)$$
which gives the Green's function
$$G_{\rm free}= {1\over \sqrt{r-v_st+i/\wedge}}\ {1\over
\sqrt{r-v_ct+{i\over\wedge}}}\qquad\qquad  (v_c=v_s)\eqno(13)$$
But we cannot assume that the connection between interacting electrons and
the modified bosons obeys (12).  The coupling of the two Fermi
surfaces which leads to the modified C.R. means that (12)
creates an object which can be thought of as a ``pseudoelectron''
with the suitable backflow caused by the fractional opposite-spin
hole which accompanies it, so it describes an exact
eigen-excitation of electron-like character moving in the exact
ground state.  These excitations are analogous to bosonized
versions of the exact eigen-excitations of charge $(I_i)$ and
spin $(J_\alpha)$ of the Lieb-Wu solution of the 1D Hubbard model.
(The discussion here was foreshadowed in Y. Ren's
thesis \cite{eleven}.) These
ladders of excitations can be described in terms of appropriate
bosons since they have linear energy-momentum relations near zero
energy, and these are the bosons which we have derived.  But the
actual electron operator creates a physical electron, not the
pseudo-electrons described by these bosons, and hence must have
the backflow compensated out.
This leads to the fractional exponents in the Green's function
and other correlation functions characteristic of the Luttinger
liquid. As in the 1D case (as shown in Ren's thesis) the
coefficients may be deduced from conservation laws and from the
Luttinger theorem of incompressibility of the Fermi sea in
momentum space.

Note that the ``pseudo-electron'' has the quantum numbers
of a true electron, and in fact it is one of the packet of exact
eigenstates created when a true electron is inserted at the
appropriate momentum, though with vanishing amplitude as $L\to
\infty$.  When a real electron is added, a cloud of particle-hole
excitations in addition to the two semions is excited, analogous
to the cloud of particle-hole excitations which causes the x-ray
edge anomaly.  This is the ``backflow''.
The modified commutation relations of the charge and
spin bosons still leave them as a bosonic description of
particles which are ``semions'' in the sense that two of them
make an electron.  The transformation which diagonalizes the C.R.
is not modified from the free particle case, i.e., it is
independent of $\eta$.

This is essentially because we maintain Luttinger's theorem of
incompressibility as a constraint, so that no net down-spin
particles are removed by the scattering process: they are merely
redistributed in momentum space, which is the ``backflow'' we
must now calculate.  $\eta\over \pi$ particles are displaced from
the neighbourhood of the scatterer particle at $k\up$, and we
must find how they displace the Fermi surface bosons, i.e., how
the phases are shifted at the Fermi surface.  But first we must
take into account some consequences of the non-Abelian spin
symmetry which we have been ignoring so far.

A key theorem of the bosonization technique follows from the
symmetry properties of the states at the Fermi surface.  As we
said before, the existence of a Fermi surface implies separate
conservation of each component of spin at each point on the Fermi
surface.  But spin conservation must remain independently of the
choice of axes, and we must be able to choose the axes at each
point independently.  A related requirement is that Kramers
degeneracy of the spin at each point of the Fermi surface
independently must be maintained.  This is not possible if spin at
different Fermi points is coupled relevantly as $\omega\to 0$.
As is seen in the 1D Hubbard model, this implies that the spin
bosons cannot acquire an anomalous dimension, and must retain the
same semionic character that they have for free Fermions.  In
our situation, this expresses itself by the observation that our
scattering calculation is slightly incomplete.  We have not
required formal spin rotation invariance (SU(2) symmetry) of the
S-matrix for scattering, which requires that the phase-shift have
the form
$$\eta=\eta_c+\eta_s(\sigma\cdot \sigma')\eqno(14)$$
and allows for a spin-flip scattering, which we have so far
ignored, of half the magnitude $\eta$ of the potential term.
This requires the scattering to take place entirely in the singlet
channel, rather than the up-down channel, as we have implied in
our discussion so far.  Our previous picture left us with one
spin $k_\up$ plus a hole of magnitude ${\eta\over \pi}$ in
$k_\down$. This left $1+{\eta\over\pi}$ $\up$ spins, but now we
have spin-flip scattering of $\eta/2$ giving ${\eta\over 2\pi}$
missing down spins and $1-{\eta\over 2\pi}$ up spins or one net
spin.  Correspondingly, this gives matching currents of up spin
in the scattered channels which leaves us with displacements only
of charge, not spin, bosons in the backflow. The co-moving hole
of magnitude $\eta/\pi$ is now in the charge channel.

In the actual 1D Hubbard model, this theorem is satisfied only to
logarithmic accuracy, leading to $(\ln \omega)^{-1}$ and $(\ln
q)^{-1}$ corrections to power laws; the relevant coupling
constant goes to zero only logarithmically.  We expect the same
pathology in 2D.
But dominant power laws will be correctly determined by
bosonization.  (All of this was foreshadowed in Haldane's
``Luttinger Liquid'' treatment of the 1D Hubbard
model.\cite{twelve})
When the spin-flip component is taken into account, we now can
determine how the phases at the Fermi
surface are shifted, specifically when we insert an electron at
$\Omega,q$ in order to calculate the one-particle Green's
function.  The rule is very simple: we calculate the phase shifts
we would have expected using naive up-spin down-spin
scattering, and replace these by phase-shifts in the pure charge
channel.  Let us first discuss the 1D case, which was worked out
by Ren \cite{eleven}.

In 1D, the amount of charge ${\eta\over\pi}$ which is displaced
from the state $k=k_F-q$ appears, half at the left-hand Fermi
point and half at the right, i.e., ${\eta\over 2\pi}$ at each.
These components multiply the Green's function by the factor
$$e^{i\theta_c^r\cdot {\eta\over 2\pi}\cdot {1\over \sqrt{2}}}\
e^{i\theta_c^\ell\cdot {\eta\over 2\pi}\cdot {1\over \sqrt{2}}}$$
which gives, in space-time representation,
a factor
$$\Big [{1\over (x^2-v^2_ct^2)}\Big ]^{({\eta\over 2\pi})^2\times{1\over
 4}}\eqno(15)$$
which, has the maximum exponent $({1\over 2})^2\times {1\over
4}={1\over 16}$, as
pointed out by Ren.
This gives the famous Fermi surface smearing exponent $2\times
1/16 =1/8$, in the strong coupling case and with the strictly
local interaction appropriate to the Hubbard model.

These two displacements are the total backflow.  The net momentum
of the backflow is zero, and the net charge $\eta/\pi$ , as it
must be.

The situation in 2D is not quite so simple.  Again, we recognize
that $\eta/\pi$ worth of charge boson---i.e., $\eta/\pi$ enclosed
by an
``internal Fermi surface''.---has been displaced from the region
of momentum $k$.  We may calculate the displacement of a circular
Fermi surface which would result from elastic incompressible deformation
of the lattice of $k$-values.
(We use a circular FS for illustrative purposes.)
This would give us
$$\vec{\delta k'}=\vec{{k'-k\over (k'-k)^2}}\cdot {\eta\over
2\pi^2}$$
and
$$\delta k_F^{(\Omega)}={\vec k_F(\Omega)\cdot \vec
{(k_F-k)}\over (k_F-k)^2} {\eta\over 2\pi^2}\eqno(16)$$
See Fig. (1).  If $k$ is chosen at $\theta =0$, and
$k=k_F-\epsilon$
$$\delta k_F(\Omega)\simeq {\epsilon\, k_F\over
\epsilon^2+k^2_F\theta^2}\cdot {\eta\over 2\pi^2} +{\eta\over
2\pi^2}\times{1-\cos\theta\over 2(1-\cos\theta)}={\eta\over
2\pi}\,\delta\,(\theta)+{\eta\over (2\pi)^2}\eqno(17)$$
In this case, half of the displacement is in the forward
direction, and half is a uniform displacement of the Fermi
level---essentially an $s$-wave, equivalent to isotropic
potential scattering.  This, however, is not quite the whole story.  In one
dimension the backflow compensated the charge and the momentum
exactly, since the left-and right-moving pieces were identical. Here,
however, we have an uncompensated momentum of the forward-moving
wave, ${\eta\over 2\pi}\times \vec k_F$.  The correct displacement
satisfying the Luttinger-Ward theorems is not merely a dilation
of the momentum lattice, but a rigid displacement of $-{\eta\over
2\pi}$ $k_F$ as well.

The simple uncompressible dilation of the Fermi surface which we
postulated in (16) is too simple: the interactions must satisfy
momentum as well as particle conservation and so the backflow
must carry no net momentum, as in 1D.  The relative $s$-wave
channel must carry momentum $-{\eta\over 2\pi} k_F$ which
compensates the extra momentum of the $\delta$-function peak at
$\vec k_F$.  This is equivalent to a uniform translation of the
Fermi surfaces, which is a simple unitary transformation
(multiplication of all states by a common factor) and does not
lead to any anomalous dimensions.  On the other hand, the
$s$-wave dilation does do so, and the anomalous dimension of the
Green's function is, as in 1D, $({\eta\over 2\pi})^2\times{1\over
2}=\alpha$, $0\le\alpha\le 1/8$.  Another way of describing this
part of the backflow is as a Fermi surface shift proportional to
$(1-2\cos\theta)$ rather than simply to 1.  This is not a
scattering in the $p$-wave channel, rather it is more like a
``Mossbauer'' zero-phonon, coherent recoil of the Fermi sea as a
whole.

The form of the Green's function is quite different from 1D: it
will look something like:
$$ G(r, t)\propto\int d\,\Omega\, e^{ik_F(\Omega)\cdot r}
\Big \{ {1\over (r\cdot \hat n(\Omega)-v_s t)^{1/2}}
{1\over
(r\cdot \hat n(\Omega)-v_c t)^{{1\over 2}+{1\over 4} ({\eta\over 2\pi})^2}}
\Big \}\times \Big ({1\over r-v_ct}\Big )^{({\eta\over
2\pi})^2\times {1\over 4}} \eqno(19)$$
$\hat n(\Omega)$ is the Fermi surface normal unit vector at
$\Omega$, and $\cos \theta =\hat n(\Omega)
 \hat r$.
Stationary phase will ensure that $G(r, t)$ comes almost
entirely from the ``patch'' $n(\Omega)||r$.

Experimentally, several hints suggest that $\alpha>1/8$ in fact,
in the cuprates.  We must not be surprised by the parallel-spin
interaction also being finite and repulsive, which will enhance
the charge-channel backflow without affecting spin properties
except to lower $v_s$ further, and make the electrons even less
Fermi-liquid.
For the Hubbard model there is a fixed relation
between $\eta_c$ and $\eta_s$ in (14), but in the physical case
$\eta$ can be larger.

Most of the physical phenomena which depend on $G$ and other
correlation functions can be calculated using the simple homogeneity
property:
$$ G={1\over t^{1+\alpha}}\ F({r\over t})\ .\eqno(20)$$
This determines the infrared spectrum in parallel and
perpendicular polarizations \cite{thirteen,fourteen},
and the Fermi surface
smearing; a similar property will give the exponent for $1/T_1$.
Only ARPES requires the full $G$, and this will depend critically
on details of the single-particle dispersion and Fermi surface,
so will require a separate investigation.

With (19) we have in principle the asymptotic solution of the 2D
electron gas with a local, repulsive interaction.  This is
expected to be valid in the regions of the phase diagram of the
Hubbard model reasonably far from half-filling (where umklapp
terms are important and can pin down the charge bosons)
and $U\to \infty$ at high density, where ferromagnetic coupling
of Landau mean field type will possibly be important, and lead to
Nagaoka ferromagnetism.  Finally, we exclude strong magnetic
fields, ``strong'' being enough to allow interference after a
full cyclotron orbit, i.e., we require $\omega_c\tau<<1$.  Under
this condition transverse gauge transformations are simple
reparametrizations of the Fermi surface and meaningless; i.e.,
the Fermi surface and anyons are mutually incompatible.
$\Omega_c\tau>>1$ destroys the symmetries implicit in the Fermi
surface and  causes gaps in the spectrum which are incompatible
with bosonization. With $\Omega_c\tau<1$ bosonization
$\rm\underline { is }$ the
gauge theory of the interacting Fermi system; there is no
meaningful other.

Khveshchenko has argued that in $\ge 2$ dimensions the equations
of motion of the bosons are a very crude approximation valid only
for very small $q$ and $\omega$.  This is clearly so in our
approach, since the ``$\delta$-function'' in equation (17) is
actually of width $q$.  We have argued that ``Chern-Simons''
types of terms are not important if $\omega_c\tau<1$, but in so
far as charge and spin velocities differ these can be effects
such as those we have postulated in the past caused by mixing of
bosons over a finite area of the Fermi surface, when electrons of
finite $q=k-k_F$ are excited.  Thus the above is a first
approximation to a much more complex theory which we do not yet
have under control.  Nonetheless it seems the only way to
proceed.

\centerline {\bf {ACKNOWLEDGEMENT}}

We would like to acknowledge valuable discussions with Y. Ren,
F.D.M. Haldane, and especially G. Baskaran; also the hospitality
of the Dept. of Theoretical Physics, Oxford.
This work was funded by the NSF under grant No. DMR 9104873.

\pagebreak

\end{document}